\documentclass[12pt,reqno,tbtags]{amsart}

\usepackage[top=3cm, bottom=2.5cm, left=1cm, right=1cm]{geometry}

\usepackage{newcent}
\usepackage[T1]{fontenc}
\usepackage[utf8]{inputenc}
\usepackage{babel}
\usepackage{bbm}
\usepackage{xspace}
\usepackage[hidelinks]{hyperref}
\usepackage{enumitem}
\usepackage[most]{tcolorbox}
\usepackage{xpatch}
\usepackage{stmaryrd}
\usepackage{float}
\usepackage{amsthm}
\usepackage{amsmath}
\usepackage{amssymb}
\usepackage{enumitem}
\usepackage{makecell}
\usepackage{graphicx}%
\usepackage{multirow}%
\usepackage{amsmath,amssymb,amsfonts}%

\usepackage{mathrsfs}%
\usepackage[title]{appendix}%
\usepackage{xcolor}%
\usepackage{textcomp}%
\usepackage{manyfoot}%
\usepackage{booktabs}%
\usepackage{algorithm}%
\usepackage{algorithmicx}%
\usepackage{algpseudocode}%
\usepackage{listings}

\usepackage{mathtools}

\usepackage{comment}
\usepackage{caption}
\usepackage{subcaption}

\usepackage{fullpage}

\newcommand{\I}{\mathbbm{1}}
\newcommand{\EXP}{\mathbb{E}}
\newcommand{\Var}{\mathbb{V}}
\newcommand{\PROB}{\mathbb{P}}

\newcommand{\inlaw}{\buildrel {\mathcal L} \over =}
\newcommand{\RR}{\mathbb{R}}
\newcommand{\GAMMA}{\mathrm{\Gamma}}
\newcommand{\isdef}{\buildrel {\rm def} \over =}

\newtheorem{lemma}{Lemma}

\newtheorem{remark}{Remark}

\usepackage[numbers, sort&compress]{natbib}

\begin{document}

\title[PearsonIV+Meixner-Morris]{Simulating random variates
from the Pearson IV and betaized Meixner-Morris distributions}

\author{Luc Devroye$\dagger$}
\thanks{$\dagger$School of Computer Science, McGill University, 
		Montr\'eal, Qu\'ebec,  Canada: {lucdevroye@gmail.com}. Supported by the Natural Sciences and Engineering Research Council of Canada (NSERC)}
 
 \author{Joe R. Hill$\ddagger$}
\thanks{$\ddagger$QBX Consulting, Austin, TX, USA: {joehill.qbx@gmail.com}}


\begin{abstract}
We develop uniformly fast random variate generators for
the Pearson IV distribution that can be used over the entire
range of both shape parameters. 
Additionally, we derive an efficient algorithm for sampling
from the betaized Meixner-Morris density, which is proportional to
the product of two generalized hyperbolic secant densities.
\end{abstract}




\maketitle
\nocite{*}

\section{Introduction}\label{intro}

The following statistical setting motivates the methods described in this paper; see Hill and Morris (2023). Let $Y_1, \ldots, Y_k$ be a sample of independent random variables,
$$
Y_i \mid \lambda \inlaw \textsc{nef-ghs}(n_i, \lambda),\, i = 1,\ldots,k,
$$
with known sample sizes $n_i \geq 1$ and common unknown parameter $\lambda \in \RR$,
where \textsc{nef-ghs} is short for the Meixner-Morris natural exponential family generated by the generalized hyperbolic secant distribution. The sampling mean and variance of $Y_i$ given $\lambda$ are
$$
\EXP \{ Y_i \mid \lambda \} = n_i \lambda
$$
and
$$
\Var \{ Y_i \mid \lambda \} = n_i (\lambda^2 + 1).
$$
Morris (1982) proved that the \textsc{nef-ghs} 
distribution is one of only six natural exponential families with quadratic variance functions (i.e., the variance is a quadratic function of the mean); the other five are the normal, Poisson, gamma, binomial, and negative binomial families. For the \textsc{nef-ghs} distribution, the variance function is $V(\lambda ) = \lambda^2 + 1$.

Let $Y_\cdot = \sum_{i = 1}^k Y_i$, $n_\cdot = \sum_{i = 1}^k n_i$, and ${\bar Y}_\cdot = Y_\cdot / n_\cdot$. The sampling distribution of $Y_\cdot$ given $\lambda$ is
$$
Y_\cdot \mid \lambda \inlaw \hbox{\textsc{nef-ghs}} (n_\cdot, \lambda),
$$
with mean and variance
$$
\EXP \{ Y_\cdot \mid \lambda \} = n_\cdot \lambda
$$
and
$$
\Var \{ Y_\cdot \mid \lambda \} = n_\cdot (\lambda^2 + 1).
$$
The statistic $Y_\cdot$ is complete sufficient for $\lambda$ and the sample average ${\bar Y}_\cdot$ is the maximum likelihood estimator and uniform minimum variance unbiased estimator of $\lambda$.

The conditional distribution of $Y_1,\ldots, Y_k$ given $Y_\cdot$ does not depend on $\lambda$. 
It is the reference distribution for checking the \textsc{nef-ghs} sampling model. The conditional means, variances, and covariances are
$$
\EXP \{ Y_i \mid Y_\cdot \} = n_i {\bar Y}_\cdot,
$$
$$
\Var \{ Y_i \mid Y_\cdot \} = \frac{ n_i (n_\cdot - n_i)({\bar Y}_\cdot^2 + 1) }{  n_\cdot + 1},
$$
and
$$
\hbox{\rm Cov} \{Y_i, Y_j \mid Y_\cdot \}
= - \frac{n_i n_j ({\bar Y}_\cdot^2 + 1) }{ n_\cdot + 1}.
$$
We call the conditional distribution of $Y_i$ given $Y_\cdot$ the betaized Meixner-Morris distribution. The name was suggested by analogy to the conditional distribution of a gamma variable given the sum of independent gamma variables being a beta distribution.

The conjugate prior for \textsc{nef-ghs} sampling is the Pearson IV family of distributions, with two parameters, a prior mean $\mu_0 \in \RR$ and a prior sample size $m_0 \geq 1$, say. The prior mean and variance are
$$
\EXP \{ \lambda \} = \mu_0
$$
and
$$
\Var \{ \lambda \}  = \frac { \mu_0^2 + 1 }{ m_0 - 1 }.
$$

The posterior distribution of $\lambda$ given $Y_\cdot$ is also a member of the Pearson IV family, with parameters $\mu_1 = (m_0 \mu_0 + Y_\cdot) / (m_0 + n_\cdot)$ and $m_1 = m_0 + n_\cdot$. The posterior mean and variance are
$$
\EXP \{ \lambda \mid Y_\cdot \} = \mu_1
$$
and
$$
\Var \{\lambda \mid Y_\cdot \} = \frac { \mu_1^2 + 1 }{ m_1 - 1}.
$$

The prior predictive marginal distribution of $Y_\cdot$ is the mixture of the \textsc{nef-ghs} sampling model averaged with respect to the Pearson IV prior distribution. The prior predictive mean and variance of $Y_\cdot$ are
$$
\EXP \{ Y_\cdot \} = n_\cdot \mu_0
$$
and
$$
\Var \{ Y_\cdot \} =  n_\cdot (\mu_0^2 + 1) \frac {m_0 + n_\cdot }{ m_0 - 1}.
$$
This distribution is the reference distribution for checking the prior distribution.

The posterior predictive marginal distribution of a future observation 
$Y_\cdot'$ is the mixture of the \textsc{nef-ghs} sampling model averaged with respect to the Pearson IV posterior distribution. 
The posterior predictive mean and variance of $Y_\cdot'$ given $Y_\cdot$ are
$$
\EXP \{ Y_\cdot' \mid Y_\cdot \} = n_\cdot \mu_1
$$
and
$$
\Var \{ Y_\cdot' \mid Y_\cdot \} = n_\cdot (\mu_1^2 + 1) \frac {m_1 + n_\cdot }{ m_1 - 1}.
$$
This distribution is used to predict future observations.

Given this setup, we need to be able to simulate random variables from the following distributions:

\begin{enumerate}
\item[(i)] the \textsc{nef-ghs} distribution; this is covered in Devroye (1993).
\item[(ii)] the betaized Meixner-Morris conditional distribution; this is covered in the second part of this note.
\item[(iii)] the Pearson IV distribution for prior and posterior samples; this is covered in the first part of this paper.
\item[(iv)] Pearson IV mixtures of \textsc{nef-ghs} densities for prior and posterior predictive marginal distributions; this can be done in two stages using (iii) followed by (i).
\end{enumerate}



\section{The Pearson IV distribution}\label{pearsonIV}

Undoubtedly, the most enigmatic member of Pearson's family of distributions (Pearson, 1895)
is the Pearson IV distribution, which is characterized by two
shape parameters, $a > 1/2$ and $s \in \RR$.  Its density on the real line
is given by
$$
f(x) = \frac {\gamma \, e^{s \arctan (x)} }{ (1+x^2)^a},
$$
where, by Legendre's duplication formula,
$$
\gamma 
\isdef  \frac { \left|  \Gamma (a - is/2) \right|^2  }{ \Gamma (a) \Gamma (a -1/2) \Gamma (1/2) }
= \frac { 4^{a-1} \left|  \Gamma (a - is/2) \right|^2  }{ \pi \Gamma (2a-1) },
$$
and $\Gamma$ is the complex gamma function.
We write $P_{a, s}$ to denote a Pearson type IV random variable with the given
parameters. 
As $P_{a, s} \inlaw - P_{a, -s}$, we will assume without loss of generality that $s \ge 0$.

The purpose of this paper is to propose random
variate generation algorithms that are uniformly fast over all choices of the parameters.
As far as we know, no explicit uniformly fast methods are known for this important distribution.
As described above in the introduction, we are personally motivated by random variates from this law being necessary in a
Bayesian framework, as the natural exponential families with quadratic variance functions derived by Morris (1982, 1983)
have Pearson conjugate families (Hill and Morris, 2023).
For example, in the notation of the introduction, we would have 
$s = m_0 \mu_0$, $a = m_0/2 +1$, and $x = \lambda$.

In Section 3, we recall some facts about $P_{a,0}$, the Student-t distribution.
In the subsequent sections, we will develop several generators
for the Pearson IV distribution. Some of these
require access to the normalization constant $\gamma$, which depends upon the complex gamma function.
One of the novelties in this paper is a simple method that does not require explicit knowledge of $\gamma$.
We recall the two design principles for the methods given below:

\begin{enumerate}
\item[(i)] The generators have to be theoretically exact; no approximation of any kind is allowed.
\item[(ii)] The expected time per random variate should be uniformly bounded over all parameter choices.
\end{enumerate}

\medskip
\noindent
In the second part of the paper, we deal with a new distribution 
whose density is proportional to the product of two densities
for the \textsc{ghs} (generalized hyperbolic secant) distribution.
Crucial for simulation in a statistical setup (Hill and Morris, 2023),
it will be called the betaized Meixner-Morris distribution.


\section{Student-t distribution}\label{Student}

The random variable $T_a$ with parameter $a > 0$ is a Student-t$(a)$ random variable if it has density
$$
\frac{1 }{ B(a/2, 1/2) \sqrt{a} ( 1+x^2 /a)^{\frac{a+1}{ 2}} },
$$
where $B$ denotes the beta function. 
First derived by Helmert (1875, 1876) and L\"uroth (1876) and later by Pearson (1895), it was named after Gosset (William S. Gosset, 1908) by Ronald Fisher.
It is in the Pearson IV family, as 
$$
\frac{ T_{2a-1} }{ \sqrt{2a-1} }  \inlaw P_{a,0}.
$$
We recall that
$$
T_a \inlaw \frac{ N }{ \sqrt{\frac{G_{a/2}}{a/2} }},
$$
where $N$ is standard normal, and $G_a$ denotes an independent gamma $(a)$ random variable.
Let $H_{a,b} = G_a/G_b$ be the ratio of two independent gamma random
variables and let $B_{a,b}$ be a beta random variable.
From the definition of the Student-t distribution,
$$
\frac{T_a^2 }{ a} = \frac{ (1/2) N^2 }{  G_{a/2} } \inlaw \frac{  G_{1/2} }{ G_{a/2} }
\inlaw  H_{1/2, a/2}
\inlaw  B_{1/2,1/2} H_{1, a/2}
\inlaw  \sin^2 (\pi U') \left( \frac{1 }{ U^{\frac{2 }{ a}} } - 1 \right),
$$
where $U$ and $U'$ are i.i.d.\ uniform $[0,1]$ random variables.
This yields a one-liner for the Student-t distribution 
due to Bailey (1994), also called the polar method for Student-t distribution:
$$
T_a \inlaw \sqrt{a}\sin (2 \pi U') \sqrt{ \frac{1 }{ U^{\frac{2}{a}} } - 1 }.
$$
See also Devroye (1996) for variations on this polar method.
Earlier methods for the Student-t distribution include algorithms by Best (1978)
and Ulrich (1984).
Very simple special cases, obtainable by the inversion method, include the Cauchy law (obtained for $a=1$), for which we have $T_1 \inlaw \tan(\pi (U-1/2))$, and the $t_2$ law, for which we have
$$
T_2 \inlaw \frac{2U-1}{\sqrt{2U(1-U)}}
$$
(see, e.g. Jones, 2002).

\section{Rejection from Student-t distribution}\label{rejection}

The obvious thing to try is to use the rejection method from Student-t distribution,
for which many uniformly fast algorithms are known.
Using
$$
f(x) \le \frac{\gamma e^{s\pi/2}}{(1+x^2)^a},
$$
it suffices to generate i.i.d.\ pairs $(X,U)$, where $X = T_{2 a -1} / \sqrt{2a-1}$ is a random variable with
density proportional to $(1+x^2)^{-a}$ and $U$ is uniform on $[0,1]$,
until 
$$
U e^{s \pi/2} \le e^{s \arctan (X)},
$$
or, equivalently, to generate i.i.d.\ pairs $(X,E)$, where $E$ is standard
exponential, until 
$$
-E + s \pi/2 \le s \arctan (X).
$$
As 
$$
{ \gamma e^{-s\pi/2} \over (1+x^2)^a} \le f(x) \le {\gamma e^{s\pi/2} \over (1+x^2)^a},
$$
it is easy to see that the expected number of iterations is 
at least $(1/2) e^{\pi s/2}$ (and at most $e^{\pi s}$),
which is uniformly bounded over all values of $a > 1/2$ and $s$ smaller than any fixed constant.
As soon as $s \ge 10$, or something of that order of magnitude, this 
simple method is unfeasible.


\section{Log-concavity}\label{logconcave}

As noted in Exercise 1 on page 308 in Devroye (1986), when $a \ge 1$, $\arctan (P_{a,s})$ has a log-concave density
on $[-\pi/2, \pi/2]$ given by
\begin{equation}
h(y) =  
\begin{cases}
\gamma e^{sy} ( \cos^2 (y) )^{a-1} & \text{if} ~|y| \le \frac{\pi }{ 2}, \\
0 & \text{else.} 
\end{cases}
\tag{1}
\end{equation}
We note that $\log (h)$ has first derivative $s -2(a-1) \tan (y)$ and 
second derivative $-2 (a-1)/\cos^2 (y)$, which is negative. 

\begin{remark}
    \textsc{the skewed cauchy family.}
When $a=1$, $h(y) = \gamma \exp(sy)$, so that a random variate with density $h$ simply is
\begin{align*}
W = 
\begin{cases}
\frac{1 }{ s} \log \left( e^{-\frac{\pi s }{ 2}} + U \left( e^{\frac{\pi s }{ 2}} - e^{-\frac{\pi s }{ 2}} \right) \right) & \text{if}~s > 0, \\
\pi (U-1/2) & \text{if}~s = 0, 
\end{cases}
\end{align*}
where $U$ is uniform on $[0,1]$. Therefore, $P_{1,s} \inlaw \tan (W)$. 
We also rediscover the standard method
for generating Cauchy random variables: $P_{1,0} \inlaw \tan (\pi (U-1/2))$.
The family of distributions  $P_{1,s}$  will be called the skewed Cauchy family. \end{remark}

In the remainder of this section, we assume that $a > 1$.
The mode of $h$ occurs at
$$
y^* = \arctan (\beta),
$$
where we set $\beta = s/(2(a-1))$. Also,
$$
h(y^*) 
= \gamma e^{s \arctan (\beta) } \left(\frac{1 }{ 1+ \beta^2} \right)^{a-1}
= \gamma \left( \frac{ e^{2 \beta \arctan (\beta)}}  { 1+ \beta^2} \right)^{a-1}.
$$
Assuming that we have constant time access to the value of $\gamma$---which requires
the complex gamma function---, we can apply the universal method for log-concave densities
from Devroye (1984), which is repeated here for the sake
of completeness.

\begin{algorithm}[H]
\caption{Log-concave method}\label{LC}
\begin{algorithmic}[1]
\State let $m$ be the location of the mode of log-concave density $h$ 
\State $M \gets h(m)$ 
\Repeat  
\State let $V$ be uniform on $[-2,2]$ 
\If {$V<-1$} 
\State replace $V$ by $-1 + \log(V+2)$ 
\ElsIf {$V>1$} 
\State replace $V$  by $1 - \log(V-1)$ 
\EndIf
\State $Y \gets m + V/M$ 
\State let $U$ be uniform on $[0,1]$ 
\Until {$U M \min ( 1 ,  \exp (1-M|Y-m|) ) \le h(Y)$} 
\State \textbf {return} $Y$ 
\Comment        {$Y$ has density $h$ }
\State {}
\Comment        {$\tan(Y) \inlaw P_{a,s}$ if $h$ is as in (1)}
\end{algorithmic}
\end{algorithm}

\begin{remark}
    \textsc{adaptive methods.}  The method given here has a uniformly bounded
time and is useful when the parameters vary in an application. For fixed parameters, several adaptive methods make the method more efficient as more random variates
are generated. See, e.g., Gilks (1992), Gilks and Wild (1992, 1993) and Gilks, Best and Tan (1995). $\square$
\end{remark}

\begin{remark}
    \textsc{references.}  Additional references on random variate generation for 
log-concave laws include H\"ormann, Leydold and Derflinger (2004), Leydold and H\"ormann (2000, 2001)
and Devroye (1986).  For an implementation of the algorithm in this section, see Heinrich (2004). $\square$
\end{remark}

\noindent
The expected number of iterations of this algorithm is four, as it is based
on the inequality
\begin{equation}
h(y) \le h(m) \min \left( 1 , e^{1- |y-m| h(m)} \right). 
\tag{2}
\end{equation}
We can apply this method here with the values $m = y^*$ and $M = h(y^*)$ given above.
As the algorithm returns $Y$ with density $h$,
$\tan (Y) \inlaw P_{a,s}$.
Note that the algorithm needs access to the exact value of $\gamma$, which
in turn depends upon the complex gamma function.


\section{Log-concavity without access to the normalization constant}\label{logconcave2}

Assume that we have a log-concave density as in (1) but have no computational access
to the normalization constant. That situation was dealt with in all generality in Devroye (2012),
but since we have a special situation here, there is a simpler solution.  The constant $\gamma$ in
(1) is easily explicitly bounded, as we will see below. Assume that 
$$
\gamma^- \le \gamma \le \gamma^+,
$$
where $\gamma^-$ and $\gamma^+$ are explicit functions of the parameters, $a$ and $s$.
In (1), we have the following value at the mode $m= y^*$ of $h$:
$$
h(m) 
=  \gamma e^{sm} ( \cos^2 (m))^{a-1} \isdef \gamma \Delta.
$$
Therefore, in our case, inequality (2) implies
\begin{equation}
h(y) \le \gamma^+ \Delta  \min \left( 1 , e^{1- |y-m| \gamma^- \Delta} \right).
\tag{3}
\end{equation}
Furthermore,
\begin{equation}
h(y) \ge \gamma^- e^{sy} ( \cos^2 (y) )^{a-1}, |y| \le \frac{\pi }{ 2}.
\tag{4}
\end{equation}
If we use the bounds (3) and (4) in the rejection method, then the expected number of iterations
is the ratio of the integrals of the bounding functions, or
$$
4 \times \frac{\gamma^+ }{ \gamma^-} \times \frac{\gamma }{ \gamma^-} \le
4 \left( \frac{ \gamma^+ }{ \gamma^-} \right)^2.
$$
As we will see in the next section, there is a nice supply of good bounds for $\gamma$ that makes
rejection based on the given inequalities uniformly fast over all values of $s$ and
all values $a \ge 1$. 
With the choice (7) given below, we see that the expected number of iterations is
$$
\le 4 \times \left( 
\frac
{ 1 + \frac{3 }{ 2 \pi^2 \sqrt{a^2 + (s/2)^2}} }
{ 1 - \frac{3 }{ 2 \pi^2 \sqrt{a^2 + (s/2)^2} } } 
\right)^4 
\times \frac{ (a+ 0.177)(a+0.677)  }{ (a+ 1/6)(a+2/3) }.
$$
Uniformly over all $a \ge 1$ and $s \ge 0$, this does not exceed
$$
4 \times \left( \frac{ 2\pi^2 +3  }{ 2 \pi^2 - 3 } \right)^4 \times \frac{ 1.177 \times 1.677 \times 18  }{ 35 } \approx 7.15.
$$
Note though that as $a \to \infty$, the expected number of iterations tends to 4, since the
inequalities for the gamma function get tighter.

Here is the complete algorithm
for generating a random variate $Y \inlaw \arctan P_{a,s}$ for $s \in \RR, a \ge 1$:

\begin{algorithm}[H]
\caption{PearsonIV generator, parameter $a \ge 1$}\label{Pearson1}
\begin{algorithmic}[1]
\State compute $\gamma^+$ and $\gamma^-$
\Comment {see (7) in the next section} 
\State $\beta \gets s/(2(a-1))$ 
\State $m \leftarrow \arctan (\beta)$ 
\Comment {the mode of log-concave density $h$}
\State $\Delta \gets e^{s m } \left(\frac{1 }{ 1+ \beta^2} \right)^{a-1}$ 
\Repeat 
\State let $V$ be uniform on $[-2,2]$ 
\If {$V<-1$} 
\State replace $V$ by $-1 + \log(V+2)$ 
\ElsIf {$V>1$} 
\State replace $V$  by $1 - \log(V-1)$ 
\EndIf
\State $Y \gets m + V/(\gamma^- \Delta)$ 
\State let $U$ be uniform on $[0,1]$ 
\Until {$|Y|\le \frac{\pi }{ 2}$ and $U \gamma^+ \Delta \min ( 1 , \exp (1-|Y-m| \gamma^- \Delta) ) \le \gamma^- e^{sY} ( \cos^2 (Y) )^{a-1} $} 
\State \textbf {return} $\tan (Y)$ 
\Comment{$\tan (Y) \inlaw P_{a,s}$ }
\end{algorithmic}
\end{algorithm}


\section{Bounding the normalization constant}\label{gammabounds}

The gamma function is defined for complex $z=x+iy$ with $x> 0$ by Euler's integral
$$
\Gamma (z) = \int_0^\infty t^{z-1} e^{-t} \, dt.
$$
Explicit inequalities for Euler's gamma function with real argument are often tied to Stirling's
approximation (Stirling, 1730).  A prime example is Robbins's upper and lower bound (Robbins, 1955).
Olver et al.\ (2023) summarize most of the well-known bounds.
Batir (2008) showed the  bounds
$$
\sqrt{2e}  \left( \frac{ x + 1/2 }{ e} \right)^{x+1/2} \le \Gamma (1+x) \le \sqrt{2\pi} \left( \frac{ x + 1/2 }{ e} \right)^{x+1/2} , x > 0,$$ 
and
$$
\sqrt{2\pi (x+ 1/6)}  \left( \frac{ x }{ e} \right)^{x}\le \Gamma (1+x) \le \sqrt{2\pi \left(x+ \frac{e^2 }{ 2\pi} -1 \right) } \left( \frac{ x }{ e} \right)^{x} , x \ge 1.
$$
Note that $e^2/(2\pi) -1 < 0.177$. It is the latter inequality that we will employ.

For $\Gamma(z)$ with $z=x+iy$, $x > 0$, 
we will use
Boyd's inequality (Boyd, 1994; see also (5.11.11) in Olver et al., 2023) which implies that
$$
\left| \Gamma (z) \right| 
= \left|\frac{ \sqrt{2 \pi }}{ z} \left( \frac{ z }{ e } \right)^{z} \right| 
\times \left| 1 + R(z) \right|,
$$
where
\begin{equation}
| R(z) | \le \frac{3 }{ 2 \pi^2 |z|}.
\tag{5}
\end{equation}
Note that
\begin{equation}
\left| \frac{ \sqrt{2 \pi }}{ z} \left( \frac{ z }{ e } \right)^{z} \right|  
= \sqrt{ \frac{ 2 \pi }{ \sqrt{x^2 + y^2} }} \left( \frac{ \sqrt{x^2 + y^2} }{ e } \right)^{x} e^{-y \arctan (y/x) }.
\tag{6}
\end{equation}
Combining Batir's last inequality with Boyd's bounds, we can derive suitable values for 
$\gamma^+$ and $\gamma^-$, 
recalling that $a \ge 1$, $s \ge 0$, and
$$
\gamma 
\isdef  \frac{ \left|  \Gamma (a - is/2) \right|^2  }{ \Gamma (a) \Gamma (a -1/2) \Gamma (1/2) }
=  \frac{ (a^2 - 1/4) a \left| \Gamma (a - is/2) \right|^2  }{ \Gamma (a+1) \Gamma (a+3/2) \sqrt{\pi} }.
$$
First we define the key constant (itself an approximation of $\gamma$),
\begin{align*}
\gamma^* 
= 
\frac{
(a-1/2) \left( 1 + (s/2a)^2  \right)^{a-1/2} e^{-s \arctan (s/2a) } 
}{
\sqrt{\pi/e} \left(  1+1/2a  \right)^a  \sqrt{a} } .
\end{align*}
Then, for the algorithm of the previous section, we suggest
\begin{align*}
\gamma^+ &= \gamma^* \times \frac{ \left (1 + \frac{3 }{ 2 \pi^2 \sqrt{a^2 + (s/2)^2 }} \right)^2  }{ \sqrt{ 1 + \frac{1 }{ 6a}}
\sqrt{ 1 + \frac{1 }{ 6(a+1/2)}} } ,\\
\gamma^- &= \gamma^* \times \frac{ \left (1 - \frac{3 }{ 2 \pi^2 \sqrt{a^2 + (s/2)^2 }} \right)^2  }
{ \sqrt{ 1 + \frac{0.177 }{ a}}  \sqrt{ 1 + \frac{0.177 }{ a+1/2}} }. \tag{7}
\end{align*}


\section{Symmetrization for parameter values $a \le 1$}\label{symmetric}

Finally, we develop a generator that is uniformly fast for $a \in (1/2,1], s \in \RR$.
As $P_{a,s} \inlaw -P_{a, -s}$, symmetrization may be helpful.
Define the symmetric density
$$
g(x) = \frac{f(x) + f(-x) }{ 2} = \frac{\gamma  \cosh ( s \arctan (x) ) }{ (1+x^2)^a }.
$$
Setting $Y=\arctan (X)$, where $X$ has density $g$ on $\RR$ yields the following 
symmetric density on $(-\pi/2, \pi/2)$:
$$
h(y) =  \gamma \cosh (sy) ( \cos^2 (y) )^{a-1}.
$$
Consider next the random variable $Z = \pi/2 - |Y|$
with density
\begin{equation}
\eta (z) =  2\gamma \cosh (s (\pi/2 -z)) ( \sin (z) )^{2(a-1)}, 0 \le z \le \pi/2.
\tag{8}
\end{equation}

For $a \in (1/2,1]$,  the density (8)
is decreasing and has an infinite peak at the origin unless $a=1$.
Most of its mass is near zero, and thus, we will attempt rejection using the bound

\begin{align*} 
\eta (z) &\le 2\gamma e^{s\pi/2} e^{-sz} (2z/\pi)^{2(a-1)}  \\
&\le 2\gamma (2/\pi)^{2(a-1)} e^{s\pi/2} e^{-sz} z^{2(a-1)} ,  0 < z \le \pi/2.
\end{align*}

We can apply rejection from the gamma distribution by generating
independent pairs $(Z,U) = (G_{2a-1}/s , U)$ until for the first time,
$Z \le \pi/2$ and
$$
U  (2Z/\pi)^{2(a-1)} \le (\sin (Z))^{2(a-1)},
$$
or, equivalently,
$$
U \le \left( \frac{ 2Z }{ \pi \sin (Z)} \right)^{2(1-a)}.
$$
The returned random variable $Z$ has density $\eta$ given in (8).
The probability of acceptance is thus at least

\begin{align*} 
\EXP &\left\{ \left( \frac{ 2G_{2a-1}/s }{ \pi \sin (G_{2a-1}/s)} \right)^{2(1-a)}  \I_{G_{2a-1}/s \le \pi/2} \right\} \\
&\ge \EXP \left\{ \frac{ 2G_{2a-1}/s }{ \pi \sin (G_{2a-1}/s)}  \I_{G_{2a-1}/s \le \pi/2} \right\} \\
&\ge \frac{2 }{ \pi}  \PROB \left\{ G_{2a-1}/s \le \pi/2 \right\}, 
\end{align*}
where $\I$ is the indicator function. By Markov's inequality, the probability in this expression is at least
$$
1 - \frac{2 \EXP \{ \frac{G_{2a-1} }{ s} \} }{  \pi} = 1 - \frac{2 (2a-1) }{ s \pi}
\ge 1 - \frac{2 }{ s \pi} 
\ge 1 - \frac{2 }{ \pi} 
$$
uniformly for all $s \ge 1, a \le 1$. In this range, the expected number of iterations of the rejection 
algorithm is at most 
$$
\frac{ \pi^2 }{ 2\pi - 4 }.
$$
Having generated $Z$ with density (8), we need to set 
$X = \tan ( S(\pi/2 -Z) )$, where $S$ is a random sign to obtain 
a random variate with the symmetrized Student-t density $g$.
Finally, a random variate with the Pearson IV density $f$ is obtained as

\begin{equation*}
\begin{cases}
X & \text{with probability} \frac{f(X) }{ f(X)+f(-X)} 
= \frac{ e^{s \arctan (X)}  }{ e^{s \arctan (X)} + e^{-s \arctan (X)}}  \\
-X & \text{else.} 
\end{cases}
\end{equation*}
We summarize the algorithm below.

\begin{algorithm}[H]
\caption{PearsonIV generator, parameter $1/2 < a \le 1$}\label{Pearson2}
\begin{algorithmic}[1]
\Repeat
\State generate $Z = G_{2a-1}/s$ and $U$ uniformly on $[0,1]$ 
\Until {$Z < \pi/2$ and $U \le \left( \frac{ 2Z }{ \pi \sin (Z)} \right)^{2(1-a)}$} 
\State $Y \gets  S(\pi/2 -Z) $, where $S$ is a random sign 
\State $X \gets \tan ( Y ) $ 
\State with probability $e^{-sY}/(e^{-sY} + e^{sY})$, $X \gets -X$ 
\State \textbf{return} $X$ 
\Comment {$X \inlaw P_{a,s}$}
\end{algorithmic}
\end{algorithm}

\begin{remark}
\textsc{gamma random variates.}
For uniformly fast gamma random variates, we refer to the surveys
in Devroye (1986) and Luengo (2022). Many simulation studies confirm
that the method of Schmeiser and Lal (1980) is quite
competitive if the gamma parameter is at least one. Xi, Tan and Liu
(2013) suggested generating $\log (G_a)$ instead. As $\log (G_a)$
has a log-concave density for all values of $a > 0$, a uniformly
fast generator is quite easily obtained either by the universal
method of Devroye (1984) or a specialized algorithm as developed, e.g.,
in Devroye (2014). $\square$
\end{remark}

\section{Putting things together for the Pearson IV distribution.}

We conclude this first part by giving an overview of methods developed above, which are all uniformly fast over the ranges of the parameters specified below:

\begin{enumerate}
\item[(i)] For $a \ge 1$ and all $s$, one can use the universal log-concave generator.
\item[(ii)] For $s \le 5$ and all $a$, one can use rejection from the Student-t density.
\item[(iii)] For $a \in (1/2, 1], s \ge 1$, one can use rejection from the gamma density after symmetrizing the Pearson distribution.
\item[(iv)] For $a = 1$, the skewed Cauchy density, there is a simple one-liner.
\item[(v)] For $s = 0$, the Student-t law, there is a simple one-liner.
\end{enumerate}


\section{The GHS distribution revisited}

Natural exponential families 
of distributions have probability density functions
of the form $e^{\theta x} \mu (dx)$
where $\mu$ is a given measure, and $\theta \in \RR$
is a natural parameter.
When we compute the mean and the variance,
and force the variance to be a quadratic function
of the mean as $\theta$ is varied,
the number of families becomes severely
restricted. Morris (1982) proved that there are in fact
only six natural exponential families with this property:
the normal, Poisson, gamma, binomial, negative binomial
and \textsc{nef--ghs} families, where \textsc{ghs} is
an abbreviation for generalized hyperbolic secant
distribution, which goes back to Harkness and Harkness (1968).
The \underbar {\textsc{nef--ghs} distribution}
with parameters $\rho > 0$ and $\lambda \in \RR$,
also called Meixner's distribution and named after
physicist Meixner (1934),
has density
\begin{equation}
f(x) = (1+ \lambda^2)^{-\rho/2} e^{x \arctan \lambda} f_\rho(x)~,
\tag{9}
\end{equation}
where $f_\rho$ is the density of the 
(\textsc{ghs}) distribution with parameter $\rho$.
While the five other families play crucial roles in
statistics, the {\textsc{nef--ghs} distribution}
became prominent as a model in the finance literature,
where it is known as Meixner's distribution.
However, Morris independently introduced and named
the {\textsc{nef--ghs} distribution} in statistics.
In the present paper, we will call it the 
\underbar{Meixner-Morris distribution}.

Schoutens and Teugels (1998) and Grigelionis (1999, 2000) first
introduced the Meixner-Morris distribution and process. Schoutens (2001, 2002, 2003) 
highlighted its uses in the financial modelling of stock markets.
Further theoretical properties were developed by Ferreira-Castilla and Schoutens (2012) and
surveyed by Mazzola and Muliere (2011).

The \textsc{ghs} density $f_\rho$ is not explicitly known in any
standard way. Its shortest description
is as a product of two gamma functions with imaginary
argument,
\begin{align*}
f_\rho (x) 
&= \frac{2^{\rho-2} }{ \pi\GAMMA(\rho)} \GAMMA \left( \frac{\rho +ix }{ 2} \right) \GAMMA \left( \frac{\rho -ix }{ 2} \right)  \\
&= \frac{2^{\rho-2} }{ \pi\GAMMA(\rho)} \left| \GAMMA \left( \frac{\rho +ix }{ 2} \right) \right|^2 \\
&= \frac{2^{\rho-2} (\GAMMA (\rho/2))^2 }{ \pi \GAMMA(\rho)} \left( \frac{\left| \GAMMA \left( \frac{\rho +ix }{ 2} \right) \right| }{ \GAMMA (\rho/2) } \right)^2 
\tag{10}
\end{align*}
(Harkness and Harkness, 1968). It is easy to see that $f_\rho$ is symmetric about $0$.

Devroye (1993) gives various uniformly fast random variate generators for
the \textsc{ghs} and Meixner-Morris distributions based on the unimodality of $f_\rho$.
For a method that requires first finding the supremum of a ratio of two functions
numerically; see Grigoletto and Provasi (2008).

Let us derive a simple representation of $f_\rho$:
\begin{equation}
f_\rho (x) = \frac{2^{\rho-2} (\GAMMA (\rho/2))^2 }{ \pi \GAMMA(\rho)} \times \EXP \left\{ \cos (xZ/2) \right\},
\tag{11}
\end{equation}
where $Z = \log (G_{\rho/2}) - \log (G'_{\rho/2})$ and $G_{\rho/2}$ and $G'_{\rho/2}$ are independent
gamma $(\rho/2)$ random variables.
It is noteworthy that both $\log (G_{\rho/2})$ and $Z$ have log-concave densities.
Still, it is not immediately clear how this would aid in random variate
generation, or even to show that $f_\rho$  is log-concave.
As we will see
in the next section, $f_\rho$ can be bounded from both sides by suitable
log-concave functions.

\noindent
\textsc{proof of (11).}
Let $G_\rho$ be a gamma $(\rho)$ random variable with $\rho>0$. 
Then $W_\rho \isdef \log (G_\rho)$ has a generalized Gumbel distribution with density
$$
\frac{e^{\rho y} \, e^{-e^y} }{ \Gamma (\rho)}, y \in \RR.
$$
Recalling the definition of the complex gamma function, we have for $\rho > 0, x \in \RR$,
$$
\Gamma (\rho+ix) = \int_0^\infty t^{\rho+ix-1} e^{-t} \, dt
$$
(see, e.g., Whittaker and Watson, 1927). Rewritten, we define the complex-valued function

\begin{align*}
h(x) 
&\isdef \frac{\Gamma (\rho+ix) }{ \Gamma(\rho) } \\
&= \int_0^\infty \frac{ t^{\rho -1} e^{-t} }{ \Gamma (\rho)} \cos (x \log t) \, dt + i \int_0^\infty \frac{ t^{\rho -1} e^{-t} }{ \Gamma (\rho)} \sin (x \log t)  \, dt \\
&= \EXP \left\{  \cos (x \log G_\rho)   + i \sin (x \log G_\rho )   \right\} \\
&= \EXP \left\{  \cos (x W_\rho)  \right\} + i \EXP \left\{ \sin (x W_\rho )  \right\}. 
\end{align*}
Define $H (x) \isdef |h(x)|^2$.
Then
\begin{align*}
H(x) 
&= \left( \EXP \left\{  \cos (x W_\rho) \right\} \right)^2 +  \left( \EXP \left\{ \sin (x W_\rho )  \right\} \right)^2 \\
&= \EXP \left\{  \cos (x W_\rho) \cos (x W'_\rho)  +  \sin (x W_\rho) \sin (x W'_\rho )  \right\}  ~\hbox{\rm (where $W_\rho$ and $W'_\rho$ are i.i.d.)}   \\
&= \EXP \left\{  \cos (x Z  )  \right\},
\end{align*}
where we define the symmetric random variable $Z \isdef W_\rho-W'_\rho$.
Now replace $x$ by $x/2$ and $\rho$ by $\rho/2$ throughout to get the desired representation. $\square$

\begin{remark}\textsc{the Meixner-Morris distribution.}
The Meixner-Morris distribution is infinitely divisible   
since its characteristic function 
has a power representation (Schoutens, 2002; see also Devianto, Safatri, Herli and Maiyastri, 2018),
$$
\varphi (t) = 
\left( \frac{ \cos (\arctan (\lambda))  }{ \cosh ( t - i \arctan (\lambda) ) } \right)^{2 \rho}.
$$
This implies, for example, that the sum of independent \textsc{ghs}
random variables with parameters $\rho_1,\ldots,\rho_n$ is
\textsc{ghs} with parameter $\sum_{i=1}^n \rho_i$.
We can associate with it a L\'evy process commonly called
the Meixner process (see, e.g., Schoutens, 2002). This L\'evy process
has no Brownian part and a pure jump part governed by the absolutely continuous 
L\'evy measure
$$
\nu (dx) = \frac{ \rho }{ x \sinh (\pi x) } \, dx
$$
for the case $\lambda = 0$ (see, Eberlein (2009), or Ferreira-Castillo and Schoutens (2012)).
\end{remark}

\section
{The betaized Meixner-Morris distribution.}

As shown in Hill and Morris (2023), and summarized in
our introduction, the conjugate prior for $\lambda$ in the
Meixner-Morris distribution is the Pearson IV distribution.
Hence it is natural to deal with both distributions at the same time.
Consider next the independent random variables $X_1,\ldots, X_n$,
where $X_i$ is Meixner-Morris distributed with
parameters $\rho_i$ and $\lambda$ in the notation of (9) and (10).
Let $S = \sum_{i=1}^n X_i$. In a statistical setting,
one would like to generate $X_1, \ldots, X_n$
given their sum $S$.  This can be done sequentially
if we know how to generate $X_1$ given the sum $S$.
The distribution of $X_1$ given $S$ does not depend on $\lambda$ and has density
$$
\frac{ f_{\rho_1} (x) f_{\sum_{i=2}^n \rho_i} (S-x) }{ f_{\sum_{i=1}^n \rho_i} (S)}, x \in \RR.
$$
For a smoother treatment, let us define the betaized Meixner-Morris density with
shape parameters $a,b > 0$ and $s \in \RR$ and denote it by $f$ in the
present section:
\begin{equation}
f(x) \isdef \frac{ f_a (x) f_b (s-x) }{ f_{a+b} (s)}, x \in \RR;
\tag{12}
\end{equation}
for example, in the notation of the introduction,
we would take $a = n_i$, $b = n_\cdot - n_i$, $x = Y_i$, and $s = Y_\cdot$.
Using (10), the density becomes

\begin{align*}
f(x) 
&= 
\frac{
\frac{2^{a-2} }{ \pi\GAMMA(a)} \left| \GAMMA(\frac{a +ix }{ 2}) \right|^2 \frac{2^{b-2} }{ \pi\GAMMA(b)} \left| \GAMMA(\frac{b +i(s-x) }{ 2})   \right|^2
}{
\frac{2^{a+b-2} }{ \pi\GAMMA(a+b)} \left| \GAMMA \left(\frac{a+b +is }{ 2} \right) \right|^2
} \\
&=
{
\frac{\Gamma (a+b) }{ 4 \pi \GAMMA(a) \GAMMA(b)} 
\frac{\left| \GAMMA \left(\frac{a +ix }{ 2}\right) \right|^2   \left| \GAMMA \left(\frac{b +i(s-x) }{ 2}\right) \right|^2
}{
\left| \GAMMA \left(\frac{a+b +is }{ 2} \right)   \right|^2 }
} . 
\end{align*}
It is well-known that the conditional distribution of a gamma variable given the sum of independent gamma variables is a beta distribution. By analogy, we call the conditional distribution of a Meixner-Morris random variable given the sum of independent Meixner-Morris variables a 
\underbar{betaized} \underbar{Meixner}\underbar{-Morris} random variable,
to avoid the acronym-based name ``betaized \textsc{ghs}''.

The mean and variance of this distribution are given by

\begin{align*}
\mu &= \frac{a }{ a+b} s, \\
\sigma^2 
&= \frac{ ab }{ (a+b)^2}  \, \frac{ s^2 +  (a+b)^2 }{ 1 + a +b } .
\tag{13}
\end{align*}
While $f$ itself is the product of two unimodal densities, the
resulting density is not necessarily unimodal.  We are
able, however, to tightly sandwich $f$ between two log-concave
functions, which will enable us to develop a uniformly fast algorithm for
sampling.

Recall from (5) and (6) that with $z=x+iy$, $r = |z| = \sqrt{x^2 + y^2}$,
$x \ge 1/2, y \in \RR$, we have
$$
\left| \Gamma (z) \right| ^2
= 
\theta \, \frac{2 \pi }{ r } \, \left( \frac{r }{ e} \right)^{2x} e^{-2y \arctan(y/x)},
$$
where $\theta$ is some number bounded as follows
$$
\left( 1 - \frac{3 }{ 2x\pi^2} \right)^2 \isdef \theta^{-}(x)  \le \theta \le \theta^{+}(x) \isdef \left( 1 + \frac{3 }{ 2x\pi^2} \right)^2.
$$
With the proper substitutions, assuming $\min (a,b) \ge 1$,
$$
f(x) = \theta g(x),
$$
where now
$$
\theta^{-}(a/2) \theta^{-}(b/2) \le \theta \le \theta^{+}(a/2) \theta^{+}(b/2),
$$
$$
g(x) 
=
\gamma
\left( \left( \frac{ a }{ 2} \right)^2 + \left( \frac{ x }{ 2} \right)^2  \right)^{\frac{a-1}{ 2}}  \,
\left( \left( \frac{ b }{ 2} \right)^2 + \left( \frac{ s-x }{ 2} \right)^2  \right)^{\frac{b-1 }{ 2}}  \,
e^{-x \arctan \left(\frac{x }{ a} \right) - (s-x) \arctan \left(\frac{s-x }{ b} \right)} ,
$$
and
$$
\gamma = 
\frac{
\Gamma (a+b) \pi }{  e^{a+b} \GAMMA(a) \GAMMA(b) \left| \GAMMA \left(\frac{a+b +is }{ 2} \right)   \right|^2 
} .
$$


\begin{lemma}
The function $g$ above is log-concave when $\min (a,b) \ge 1$.
\end{lemma}

\noindent
\textsc{proof.}
As $g$ consists of a product of two functions, it suffices to establish
the log-concavity of each, to conclude that $g$ is log-concave.
So, let us prove that
$$
( (a/2)^2 + (x/2)^2  )^{(a-1)/2}  \,
e^{-x \arctan \left( \frac{ x }{ a} \right)},
$$
is log-concave, or, equivalently, that
$$
( a^2 + x^2  )^{a-1}  \, e^{-2x \arctan \left( \frac{ x }{ a} \right)}
$$
is log-concave. 
Define the logarithm of the latter function as
$$
h(x) = (a-1) \log \left( a^2 + x^2 \right) - 2 x \arctan \left( \frac{ x }{ a} \right).
$$
We have
$$
h'(x)
=
\frac{2(a-1) x  }{  a^2 + x^2 }  - 2 \arctan \left( \frac{ x }{ a} \right) - \frac{ 2x }{ a (1 + \left( \frac{ x }{ a} \right)^2 ) }
=
\frac{- 2x  }{  a^2 + x^2 }  - 2 \arctan \left( \frac{ x }{ a} \right) .
$$
Thus,

\begin{align*}
h''(x)
&=
- \frac{ 2  }{  a^2 + x^2}  + \frac{ 4x^2 }{ (a^2 + x^2)^2 } - \frac{2 }{ a (1 + \left( \frac{ x }{ a} \right)^2 ) } \\
&=
\frac{ -2(a^2 + x^2)+4x^2 -2a(a^2 + x^2)   }{  (a^2 + x^2)^2}   \\
&=
\frac{ -2a^2 + 2x^2 -2a^3 -2ax^2   }{  (a^2 + x^2)^2}   \\
&=
- \frac{ 2(a-1)x^2 + 2 (a+1) a^2 }{(a^2 + x^2)^2}  ,
\end{align*}
which is nonpositive when $a \ge 1$.  $\square$
\medskip

In summary, we have a random variate generation problem on our hands
where the target density $f$ can be bounded as follows:
$$
\alpha g(x) \le f(x) \le \beta g(x),
$$
where $g$ is proportional to a log-concave density,
\begin{equation}
\alpha = (1-3/(a\pi^2))^2 (1-3/(b\pi^2))^2 ,
\tag{14}
\end{equation}
and
\begin{equation}
\beta = (1+3/(a\pi^2))^2 (1+3/(b\pi^2))^2.
\tag{15}
\end{equation}
We do not know the mode and mean of that density
proportional to $g$, but are given the mean
and variance of the target density $f$.  
This situation leads to the algorithms developed below.
The betaized Meixner-Morris distribution is but a special case.
We have the following inequality:


\begin{lemma}
The following inequality holds when $\min(a,b) \ge 1$:
$$
f(x) \le \frac{A }{ \sigma} \min \left( 1 , e^{B - \frac{C |x-\mu| }{ \sigma}} \right), x \in \RR,
$$
where $\mu$ and $\sigma^2$ are the mean and variance of $f$ (see (13)),
$$
A = \beta^2 ,
$$
$$
B = 1+ \frac{ 1 + \sqrt{ 3(1 + 1/\alpha^2) }  }{ \sqrt{12 + 12/\alpha^2}   }
= \frac{3 }{ 2} + \frac{1 }{ \sqrt{12 + 12/\alpha^2} },
$$
$$
C = \frac{ \alpha }{ \sqrt{12 + 12/\alpha^2}},
$$
and $\alpha$ and $\beta$ are as in (14) and (15).
\end{lemma}

\noindent
\textsc{proof.}
Let us introduce the notation
$c_g$, $m_g$, $M_g$, $\sigma^2_g$ and $\mu_g$ for
the integral of $g$, mode of $g$, the value of the density $g/c_g$ at that mode,
the variance of $g/c_g$ and the mean of $g/c_g$.
Without subscripts, these quantities would be 
$1, m, M, \sigma^2$ and $\mu$ for the target density $f$.
We begin with the inequality
$$
f(x) \le  \beta g(x) = {\beta c_g } \frac{g(x) }{ c_g} 
\le \beta c_g \, M_g \min \left( 1 , e^{1- c_g M_g |x-m_g| } \right), x \in \RR.
$$
We will upper bound $c_g M_g$ and $|m_g - \mu_g|$ and
lower bound $c_g M_g$ to prove the Lemma.
First of all, the Johnson-Rogers inequality (1951) for
unimodal densities implies that
$$
| m_g - \mu_g | \le \sqrt{3} \sigma_g;
$$
see Dharmadhikari and Joag-Dev (1988) and Bottomley (2004).
By taking integrals, we also have
$$
\alpha \le c_g \le \beta.
$$
To relate $\mu$ and $\mu_g$, we note that
$$
| \mu - \mu_g  |
= \left| \int (x - \mu) \frac{g }{ c_g} \right|
\le \sqrt{ \int (x - \mu)^2 \frac{g }{ c_g} } 
\le \sqrt{ \int (x - \mu)^2 \frac{f }{ \alpha c_g} } 
\le \sqrt{ \int (x - \mu)^2 \frac{f }{ \alpha^2 } } 
= \frac{\sigma }{ \alpha } .
$$
Furthermore,
\begin{align*}
\sigma^2_g 
&=  \int (x - \mu_g)^2 \frac{g }{ c_g } \\
&\le  \int (x - \mu_g)^2 \frac{f }{ c_g \alpha}  \\
&\le  \int (x - \mu_g)^2 \frac{f }{ \alpha^2}  \\
&=  \int (x - \mu)^2 \frac{f }{ \alpha^2}  + \frac{(\mu-\mu_g)^2 }{ \alpha^2} \\
&= \frac{ \sigma^2 + (\mu-\mu_g)^2 }{ \alpha^2} \\
&\le \frac{ \sigma^2 (1 + 1/\alpha^2)  }{ \alpha^2}. 
\end{align*}
Combining this and using the triangle inequality, we see that
$$
| m_g - \mu | 
\le \frac{ \sqrt{3}\sigma }{ \alpha} \sqrt{ 1 + 1/\alpha^2 } + \frac{ \sigma }{ \alpha} 
= \frac{ \sigma }{ \alpha} \left( 1 + \sqrt{ 3(1 + 1/\alpha^2) } \right) .
$$
Furthermore,
\begin{align*}
\sigma^2_g 
=  \int (x - \mu_g)^2 \frac{g }{ c_g } 
\ge  \int (x - \mu_g)^2 \frac{f }{ c_g \beta}  
\ge  \int (x - \mu)^2 \frac{f }{ \beta^2}  
= \frac{ \sigma^2 }{ \beta^2}. 
\end{align*}

By Bobkov and Ledoux (2019, proposition B2),
we have for any log-concave density $h$ with mode $m$
and variance $\sigma^2$,
$$
\frac{1 }{ \sqrt{12}} \le \sigma h(m)  \le 1.
$$
The left-hand side of this inequality
is valid for all densities, not just the log-concave 
densities (see Statulevicius (1965) and Hensley (1980)).
The right-hand side of the inequality is 
due to Fradelizi, Gu\'edon and Pajor (2014; see also Bobkov and Chistyakov, 2015).
Thus,
$$
\frac{ \alpha }{ \sigma\,\sqrt{12 + 12/\alpha^2} }
\le
\frac{ 1 }{ \sqrt{12} \, \sigma_g }
\le
c_g M_g 
\le \frac{1 }{ \sigma_g} 
\le \frac{\beta }{ \sigma }.
$$
Therefore,
\begin{align*}
f(x) 
&\le \beta c_g \, M_g \min \left( 1 , e^{1- c_g M_g |x-m_g| } \right) \\
&\le \frac{ \beta^2 }{ \sigma} \min \left( 1 , e^{1+ | m_g - \mu |\frac{ \alpha }{ \sigma \sqrt{12 + 12/\alpha^2}  } - \frac{ \alpha }{ \sigma \sqrt{12 + 12/\alpha^2} } |x-\mu| } \right) \\
&\le \frac{ \beta^2 }{ \sigma} \min \left( 1 , e^{1+ \frac{ 1 + \sqrt{ 3(1 + 1/\alpha^2) }  }{ \sqrt{12 + 12/\alpha^2}   } - \frac{ \alpha }{ \sigma \sqrt{12 + 12/\alpha^2} } |x-\mu| } \right). 
\end{align*}
The proof follows by inspection of all the coefficients. $\square$


The bound on $f$ consists of a constant piece of value $A/\sigma$ centered at $\mu$
of width $2B\sigma/C$, and two exponential tails, to the right of $\mu + B\sigma/C$ and 
to the left of $\mu - B\sigma/C$.  The total area under the constant piece is $2AB/C$
and under both exponential tails, $2A/C$.  Therefore, a rejection method based on
Lemma 2, has an expected number of iterations equal to
$$
2 (1+B) \frac{A }{ C},
$$
which only depends upon $\alpha$ and $\beta$, and thus remains bounded uniformly over
all parameter choices $a,b \ge 1$. As $a, b \to \infty$, we see that $\alpha, \beta \to 1$,
$A \to 1$, $B \to 1+ (1+\sqrt{6})/\sqrt{24}$, $C \to 1/\sqrt{24}$, so that
the expected number of iterations tends to
$$
2 ( 2\sqrt{24} + 1 + \sqrt{6}) = 26.49\ldots .
$$
Recall that these are uniform bounds of a one-size-fits-all generic algorithm.
We now have all the ingredients for the first algorithm for the
betaized Meixner-Morris distribution.

\begin{algorithm}[H]
\caption{Betaized Meixner-Morris generator, parameters $a,b \ge 1$}\label{BetaMeixner}
\begin{algorithmic}[1]
\State $\alpha \gets (1-3/(a\pi^2))^2 (1-3/(b\pi^2))^2$ 
\State $\beta \gets (1+3/(a\pi^2))^2 (1+3/(b\pi^2))^2$
\State $A \gets \beta^2$ 
\State $B \gets \frac{3 }{ 2} + \frac{1 }{ \sqrt{12 + 12/\alpha^2} }$ 
\State $C \gets \frac{ \alpha }{ \sqrt{12 + 12/\alpha^2}}$ 
\State $\mu \gets \frac{a }{ a+b} s$
\State $\sigma \gets \sqrt{\frac{ ab }{ (a+b)^2}  \, \frac{ s^2 +  (a+b)^2 }{ 1 + a +b }}$ 
\Repeat 
\State let $V$ be uniform on $[0,1]$ 
\If {$V \le B/(1+B)$}
   \State $X \gets \mu+ B V'  \sigma/C$ where $V'$ is uniform on $[-1,1]$ 
\Else 
   \State $X \gets \mu+ S(B+E)\sigma/C$ where $E$ is exponential, 
   \State \phantom{00000} and $S$ is a random sign 
\EndIf
   \State let $U$ be uniform on $[0,1]$ 
\Until {$U (A/\sigma) \min ( 1 ,  \exp (B-C|X-\mu|/\sigma) ) \le f(X)$} 
\State \textbf{return} $X$ 
\Comment{$X$ has density $f$}
\end{algorithmic}
\end{algorithm}

We can tighten the inequality of Lemma 2 by a more delicate argument, resulting
in the following Lemma.


\begin{lemma}
Let $\mu$ and $\sigma^2$ be the mean and variance of $f$ (see (13)).
Let $\alpha$ and $\beta$ be as in (14) and (15).
Define
$$
\eta = \frac{ \sigma }{ \alpha} \left( 1 + \sqrt{ 3(1 + 1/\alpha^2)} \right),
$$
$$
\tau = \frac{ \alpha }{ \sigma \, \sqrt{12 + 12/\alpha^2}  },
$$
and
$$
\tau' = \frac{\beta }{ \sigma }.
$$
The following inequality holds when $\min(a,b) \ge 1$:
$$
f(x) 
\le 
\begin{cases}
\beta \tau' & \text{if} ~|x-\mu| \le \eta + \frac{1 }{ \tau'}  \\
\frac{ \beta }{ |x-\mu|-\eta } & \text{if} ~\eta + \frac{1 }{ \tau'} \le |x-\mu| \le \eta + \frac{1 }{ \tau}  \\
\beta \tau \min \left( 1 , e^{1+ \eta \tau  -\tau |x-\mu| } \right) & \text{if} ~|x-\mu| \ge \eta + \frac{1 }{ \tau}. 
\end{cases}
$$
\end{lemma}

\noindent
\textsc{proof.}
We keep the notation of the proof of Lemma 2.
We begin with inequality
$$
f(x) \le  \beta g(x) = {\beta c_g } \frac{g(x) }{ c_g} 
\le \beta c_g \, M_g \min \left( 1 , e^{1- c_g M_g |x-m_g| } \right), x \in \RR.
$$
Recalling
$$
| m_g - \mu | 
\le \frac{ \sigma }{ \alpha} \left( 1 + \sqrt{ 3(1 + 1/\alpha^2) } \right) ,
$$
and
$$
\tau \isdef \frac{ \alpha }{ \sqrt{12 + 12/\alpha^2}  \sigma }
\le
c_g M_g 
\le \frac{\beta }{ \sigma }
\isdef \tau',
$$
we obtain
$$
f(x) 
\le \beta c_g \, M_g \min \left( 1 , e^{1+ \eta c_g M_g  -c_g M_g |x-\mu| } \right)
\le \max_ {\tau \le t \le \tau'} \beta t \min \left( 1 , e^{1+ \eta t  -t |x-\mu| } \right) .
$$
For $|x-\mu| \le \eta$, the best we can hope for is the bound $f(x) \le \beta \tau'$.
Assume next that $|x-\mu| > \eta$. Then define the following threshold
$t^* = 1/(|x-\mu|-\eta)$, which makes the exponent in the upper bound zero.
If $\tau' \le t^*$, then $f(x) \le \beta \tau'$.
If $\tau \ge t^*$, then 
$$
f(x) \le \beta \tau \min \left( 1 , e^{1+ \eta \tau  - \tau |x-\mu| } \right).
$$
Finally, if $\tau \le t^* \le \tau'$, we bound by
$$
f(x) \le \beta t^* = \frac{ \beta }{ |x-\mu|-\eta }.
$$
Combining this yields the bound
$$
f(x) 
\le 
\begin{cases}
\beta \tau' & \text{if}~ \tau' \le t^* = \frac{ 1 }{ |x-\mu|-\eta}  \\
\beta t^* = \frac{ \beta }{ |x-\mu|-\eta } & \text{if}~ \tau \le t^* \le \tau'   \\
\beta \tau \min \left( 1 , e^{1+ \eta \tau  -\tau |x-\mu| } \right) & \text{if} ~\tau \ge \frac{ 1 }{ |x-\mu|-\eta}. 
\end{cases}
$$
This translates into the bound given in Lemma 3.
$\square$


We can develop a slightly more complex rejection algorithm as we have a tripartite 
dominating function.  
The only novelty is the central portion of the upper bound.
In this respect, we observe that a random variate with density proportional to $1/x$
on an interval $[x', x'']$ can be generated as $1/(x''^{1-U} x'^U )$,
where $U$ is uniform on $[0,1]$.
We begin with the calculation of the area under the dominating curve, which is,
for the central part,
$$
2 \beta \tau' \left( \eta + \frac{1 }{ \tau'} \right),
$$
for the middle portion,
$$
2 \beta \log \left( \frac{ \tau' }{ \tau } \right),
$$
and for the exponential tails,
$$
2 \beta.
$$
Again, the sum is a continuous function of $\alpha$ and $\beta$, and thus
uniformly bounded away from infinity over all choices of $a,b \ge 1$ and $s \in \RR$.
For comparison with the bound of Lemma 2, we compute the limiting value of the area
as $a,b \to \infty$ as
$$
2  ( 3 + \sqrt{6} + (1/2) \log (24) ) = 14.07\ldots .
$$
Thus, Lemma 3 increases the efficiency by almost 60\% for large values of the parameters.
The algorithm is given below.

\begin{algorithm}[H]
\caption{Betaized Meixner-Morris generator, parameters $a,b \ge 1$}\label{BetaMeixner2}
\begin{algorithmic}[1]
\State $\mu \gets \frac{a }{ a+b} s$ 
\State $\sigma \gets \sqrt{\frac{ ab }{ (a+b)^2}  \, \frac{ s^2 +  (a+b)^2 }{ 1 + a +b }}$ 
\State $\alpha \gets (1-3/(a\pi^2))^2 (1-3/(b\pi^2))^2$ 
\State $\beta \gets (1+3/(a\pi^2))^2 (1+3/(b\pi^2))^2$ 
\State $\eta \gets \frac{ \sigma }{ \alpha} \left( 1 + \sqrt{ 3(1 + 1/\alpha^2)} \right)$ 
\State $\tau \gets \frac{ \alpha }{ \sigma \, \sqrt{12 + 12/\alpha^2}  }$
\State $\tau' \gets \frac{\beta }{ \sigma }$ 
\State $q_1 \gets \beta (1+ \tau' \eta) $  
\State $q_2 \gets \beta \log (\tau' / \tau)$ 
\State $q_3 \gets \beta $ 
\State $q = q_1 + q_2 + q_3 $ 
\Repeat 
\State let $U,V$ be uniform on $[0,1]$ 
\If {$V \le q_1/q$ }
 \State $X \gets \mu+ V'(\eta + 1/\tau')$ where $V'$ is uniform on $[-1,1]$ 
\State Accept $\gets [ U \beta \tau' \le f(X) ]$ 
\ElsIf {$V \le q_2/q$} 
\State $Y \gets 1/({\tau'}^{1-V'} \tau^{V'})$ where $V'$ is uniform on $[0,1]$ 
\State $X \gets \mu+ S( \eta + Y )$ where $S$ is a random sign 
\State Accept $\gets [ U \beta / Y \le f(X) ]$ 
\Else 
\State $X \gets \mu + S( \eta + (1+E)/\tau )$ where $E$ is exponential
\State \phantom{00000} and $S$ is a random sign 
\State Accept $\gets [ U \beta \tau e^{-E} \le f(X) ]$ 
\EndIf
\Until {Accept} 
\State \textbf{return} $X$ 
\Comment{$X$ has density $f$} 
\end{algorithmic}
\end{algorithm}

\section
{Conclusion.}

We presented uniformly fast algorithms for the Pearson IV distribution over the entire
range of parameters. We also introduced the betaized Meixner-Morris distribution and developed
uniformly fast algorithms for that three-parameter family, provided that the first two
parameters ($a$ and $b$) have values at least equal to one.

\bibliographystyle{plainnat}
\bibliography{p.bib}
\end{document}